\documentclass[proceedings]{rmaa}

\usepackage{rmaacite}

\renewcommand{\P}[1]{%
\ifnum#1=1\hbox{OW~168--326E}\fi
\ifnum#1=2\hbox{OW~167--317}\fi
\ifnum#1=3\hbox{OW~163--317}\fi
\ifnum#1=5\hbox{OW~158--323}\fi
\ifnum#1=0\hbox{OW~171--334}\fi}

\title{The Age of the Galaxy from Thorium Cosmochronometry}
\author{C.~Sneden\altaffilmark{1} and J.~J.~Cowan\altaffilmark{2}}
\altaffiltext{1}{University of Texas at Austin}
\altaffiltext{2}{University of Oklahoma}

\fulladdresses{
\item C. Sneden: Department of Astronomy and McDonald Observatory,
University of Texas, Austin, TX 78712, USA
\item J. J. Cowan: Department of Physics and Astronomy, University of 
Oklahoma, Norman, OK 73019}

\shortauthor{Sneden \& Cowan}
\shorttitle{Thorium Cosmochronometry}


\abstract{
The long-lived radioactive element thorium can potentially 
provide a simple and direct clock to determine the age of our Galaxy.
Spectroscopic investigations of thorium in metal-poor stars
have yielded some promising initial results.
We discuss the major observational and theoretical limitations
in thorium cosmochronometry, and point out the ways in which
the implied Galactic ages from thorium abundances can be made more accurate.
}

\listofauthors{C.~Sneden \& J.~J.~Cowan}
\indexauthor{Sneden, C.}
\indexauthor{Cowan, J.~J.}

\begin{document}

\maketitle

\section{Introduction}

Determination of the age of the universe has been a fundamental goal
of astronomical research for centuries.
Modern age-dating methods of varying degrees of directness attack
both extragalactic objects (e.g., galaxies of all types, via the Hubble Law), 
and the older galactic objects (e.g., globular cluster main sequence
turnoff stars, white dwarf stars).
Every attempt to transform the observed properties of astronomical objects 
into their ages involves some estimates and assumptions; each 
of these leaps of faith inevitably increases the total derived age 
uncertainties.

Nuclear cosmochronometry begins with the determination of relative abundances 
of stable and long-lived radioactively unstable very heavy elements.
The abundance ratios are used to determine the time since
the synthesis of these elements, given that the half-lives of the
unstable elements are known from laboratory studies.
The elements involved in nuclear cosmochronometry are all 
neutron-capture (or n-capture) elements. 
They are in the atomic number domain Z~$>$~30, and their isotopes are 
overwhelmingly created via bombardment of lighter target nuclei by free
neutrons.
The bombardment can be slow enough that almost any possible $\beta$-decays 
of too-neutron-rich nuclei has time enough to occur between successive 
neutron captures.
This so-called ``{\it s}-process'' synthesis commonly occurs in helium fusion 
zones during the late quiescent stages of stellar evolution, as evidenced 
by the n-capture-rich asymptotic giant branch carbon stars.
At the other neutron bombardment extreme, an extremely large neutron
flux overwhelms the $\beta$-decay rates, pushing nuclei out to the
``neutron drip line'' in a matter of seconds, which then decay back
toward the valley of $\beta$-stability after the neutron blast shuts off.
The site of this ``{\it r}-process'' synthesis is not known with certainty,
other than the realization that it must be at or near the point of stellar
death; low-mass supernovae, neutron-star binaries, and other massive star
sites have all been suggested (see Cowan et al. 1999 for a more detailed 
discussion and references).

Thorium (Z~=~90) and uranium (Z~=~92) are radioactive n-capture 
elements with at least three properties in common. 
First, they can be synthesized only in the ${\it r}$-process, because the
heaviest completely stable element is bismuth (Z~=~83).  
Elements in the atomic number domain 84~$\leq$~Z~$\leq$~89 radioactively 
decay so quickly that {\it s}-process synthesis occurs too slowly 
to be able to bridge this six-element gap.
Second, thorium and uranium decay on giga-year time scales, slowly enough 
that they can be used as chronometer elements for the solar system and the
Galaxy.
Third, these elements are spectroscopically detectable in the Sun and
stars.
For all of these reasons thorium and uranium will be the focus of the rest of
the present paper.
We first briefly review the history of thorium abundance studies in
stars, with emphasis on the unique n-capture-rich very metal-poor 
star CS~22892-052 (\S2), and then discuss the major uncertainties
in thorium and uranium cosmochronometry, and how these uncertainties may
be reduced with future work (\S3).

\section{Past and Present Thorium Abundance Studies}

The first extensive study of stellar thorium cosmochronometry was published by
Butcher (1987), who synthesized the 4019~\AA\ line of \ion{Th}{2}, and
compared the derived thorium abundances to those of neodymium derived from
a neighboring \ion{Nd}{2} line.
This analysis was refined and extended by Morell et al. (1992).
One significant limitation of these pioneering studies was the use
of neodymium as a comparison element: it is made in solar-system material
in roughly equal measure by the {\it s}- and {\it r}-process (Burris
et al. 2000, and references therein), but the
fractional contributions to its abundance in the target stars was
not assessed.
Additionally, these studies only considered relatively metal-rich
disk stars, whose ages compared with the oldest halo stars cannot
easily be determined.
A major step forward was achieved by Fran\c{c}ois et al. (1993),
whose thorium abundance study targeted very metal-poor halo field giants,
and used europium (a 97\% {\it r}-process element in solar-system material;
Burris et al.) as a more appropriate comparison stable element.
That investigation found a large star-to-star scatter in derived
[Th/Eu]\footnote{
[A/B]~$\equiv$~log$_{\rm 10}$(N$_{\rm A}$/N$_{\rm B}$)$_{\rm star}$~--
log$_{\rm10}$(N$_{\rm A}$/N$_{\rm B}$)$_{\odot}$}
ratios, but there were heterogeneous sources for the europium
abundances, and no definitive statements could be made sources of
the [Th/Eu] scatter.
They noted that at the time of their paper, ``The complex variation of the
Th/Eu ratio weakens the use of this ratio as a radiochronometer as long as no
detailed yields will be available for these elements'' (see \S3.2).

\begin{figure}
  \begin{center}
    \leavevmode
    \includegraphics[width=13cm]{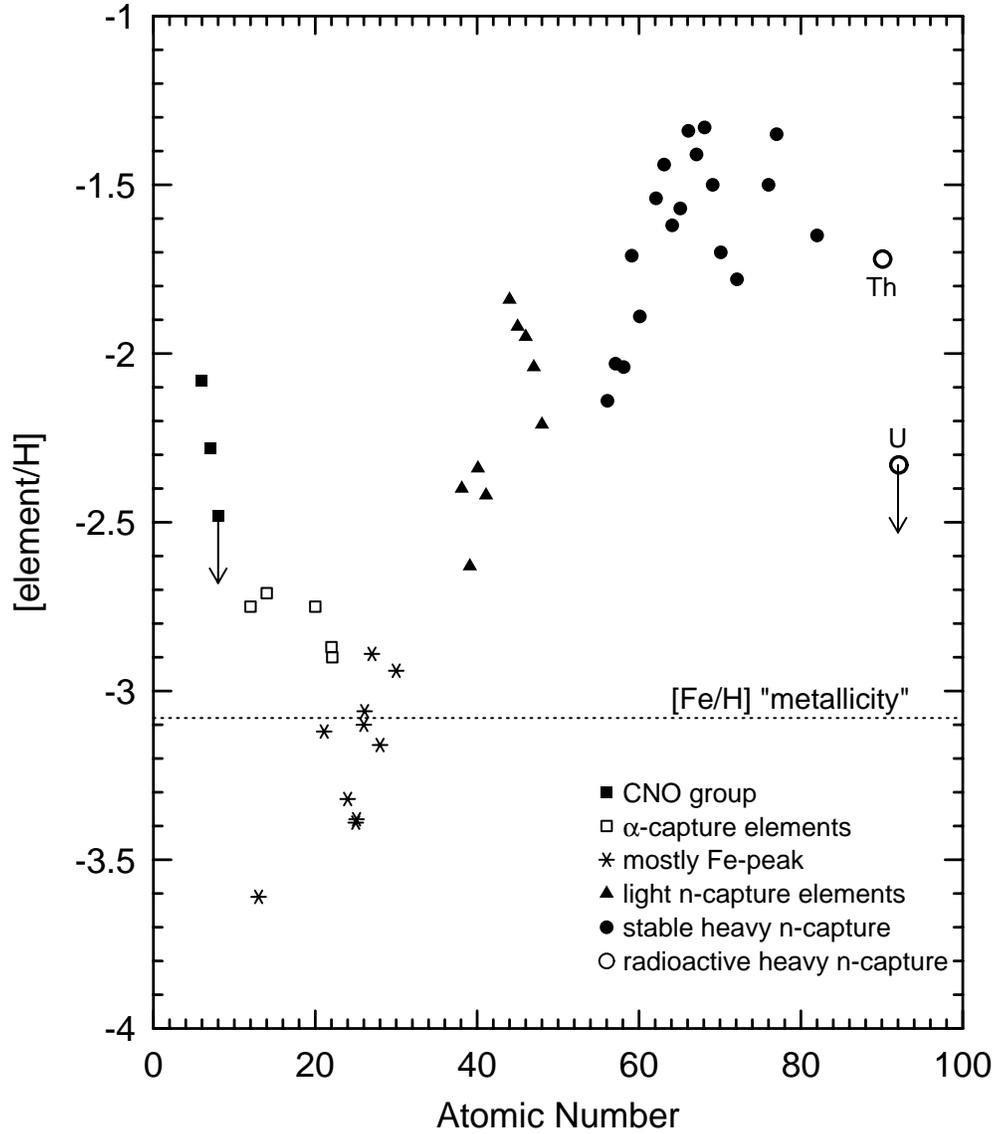}
    \caption{Abundances of 46 elements in the very metal-poor giant star
      CS~22892-052.  Note the extreme overabundances of the
      n-capture elements, which grows with increasing atomic
      number with the conspicuous exception of the radioactive
      elements thorium and uranium. See Sneden et al. (1996,2000)
      and Cowan et al. (1999) for additional discussion of the
      elemental abundances not discussed here.}
    \label{aball}
  \end{center}
\end{figure}

Included as a target star in a large spectroscopic survey of abundances in 
newly-discovered very metal-poor giants (McWilliam et al. 1995), the 
star CS~22892-052 was serendipitously discovered to have extremely
large relative overabundances of all detectable n-capture elements,
and the abundance pattern among these elements was consistent
with a scaled solar-system {\it r}-process pattern (Sneden et al. 1994).
The spectrum of the star was subjected to further scrutiny with
better spectra (more extensive wavelength coverage, higher S/N
and resolution) by Sneden
et al. (1996), and  Norris, Ryan, \& Beers (1997a), and recently
Sneden et al. (2000) have published a new abundance study that includes
about half of the elements in the periodic table.
In Figure~\ref{aball} we summarize the results of that work.
The abundances of CS~22892-052 clearly exhibit a non-solar pattern.
The parts of the abundance distribution relevant to the present 
discussion are: {\it (a)} the very low metallicity ([Fe/H]~=~--3.1)
which ensures that the abundances of CS~22892-052 were created near
the nucleosynthetic beginning of the Galaxy; {\it (b)} the extreme
overabundances of most n-capture elements (roughly peaking at
[Dy/Fe]~$\simeq$~+1.8),
suggesting that a single nucleosynthesis event was responsible for the
creation of this star's n-capture elements; {\it (c)} the steady increase
in overabundances from barium to europium, and the flat abundance 
distribution of the stable n-capture elements beyond europium, indicating 
a dominant {\it r}-process synthesis mechanism for 
(at least) the heaviest n-capture elements; and {\it (d)} the relatively
lower abundance of thorium and the even lower abundance upper limit for uranium.
This latest study reaffirms that for elements with Z~$\geq$~56, there
is an almost perfect abundance match between the observed CS~22892-052
and a scaled solar-system {\it r}-process-only abundance distribution.

CS~22892-052 has the largest set of n-capture elemental overabundances 
with a distinct {\it r}-process signature ever found in a metal-poor
halo star.
But it stands alone only in the magnitude of the n-capture overabundances, 
not in the {\it r}-process nature of these abundances.
For example, a recent detailed study (Westin et al. 2000) of the very 
metal-poor giant star HD~115444 reveals a nearly identical n-capture 
abundance pattern.  
More generally, Burris et al. (2000) show that the dominance of the
{\it r}-process is very general among halo stars with [Fe/H]~$\lesssim$~--2.5.
The interpretation of {\it r}-process synthesis of n-capture elements
in the early Galactic halo seems reasonably secure. 
The apparently uniform production of the heavier {\it r}-process
elements throughout the Galaxy's history is not well understood, but
with this fairly uniform observational result we can now turn attention 
to observation and interpretation of thorium abundances.

\section{Thorium Abundances and Cosmochronometry}

\subsection{Observational Aspects of Thorium and Uranium}

The ionization potential of \ion{Th}{1} is low, and so thorium 
is essentially totally ionized in stellar atmospheres.
There are many lines in the \ion{Th}{2} spectrum, but the relatively low 
thorium abundance renders most of the transitions undetectably
weak in low metallicity stars.
Until recently, the only \ion{Th}{2} line studied in cool stars
has been the 4019.12~\AA\ feature.
Analysis of this line is very difficult for a number of reasons.
First and foremost, it is severely blended with a number of
atomic and molecular ($^{\rm 13}$CH) features (e.g., Morell et al. 1992, 
Fran\c{c}ois et al. 1993, Norris et al. 1997b).
The blending agents are of comparable strength or greater than
that of the \ion{Th}{2} line in metal-poor giants with [Fe/H]~$\lesssim$~0.0 
(and they totally dominate the thorium feature in higher metallicity 
disk main sequence stars; see Figure~1 of Morell et al. 1992), so this 
feature is useless for reliable age estimates unless [Th/Eu]~$\gtrsim$~+0.5 
(e.g., see Figure~7 of Westin et al. 2000).

\begin{figure}
  \begin{center}
    \leavevmode
    \includegraphics[width=13cm]{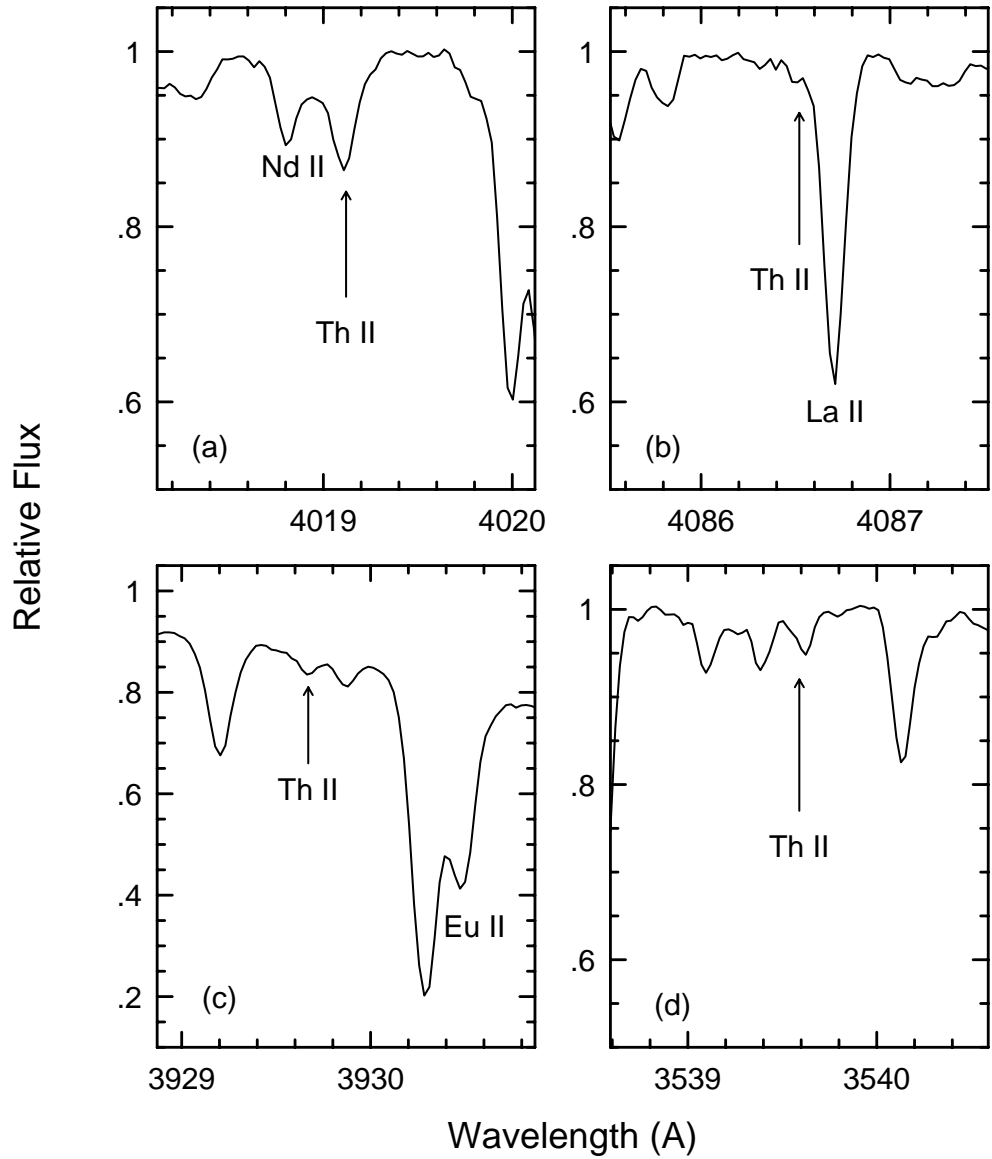}
    \caption{Small spectral regions surrounding four of the
      strongest \ion{Th}{2} lines.  Panel (a) displays the
      4019~\AA\ feature used exclusively in past thorium abundance
      studies, panel (b) shows the newly-detected 4086~\AA\
      line, and panels (c) and (d) have the spectra of two
      proposed new \ion{Th}{2} lines. Attribution of the
      features in panels (c) and (d) solely
      or mostly to thorium is tentative at present.}
    \label{thspectra}
  \end{center}
\end{figure}

One obvious way to increase the reliability of thorium abundances
would simply be to detect other \ion{Th}{2} transitions in metal-poor
stars.
But all other lines have smaller transition probabilities than the 
already-weak 4019~\AA\ line, so the task is not easy.
The n-capture-rich star CS~22892-052 is the obvious target to
search for secondary \ion{Th}{2} lines. 
Sneden et al. (2000) report detection of a line at 4086.52~\AA, suggesting 
that the thorium abundance derived from this feature is in good agreement
with the abundance from the 4019~\AA\ line.
In Figure~\ref{thspectra} we show these two lines, as well as two additional
ones that lie at the wavelengths of predicted \ion{Th}{2} lines,
appear to be largely free of blending, and have strengths roughly
consistent with ones expected from extrapolation of the strengths
of the 4019 and 4086~\AA\ lines.
Further investigation is needed before these suggested new thorium
transitions can yield reliable abundances.
Unfortunately, it is clear from the spectra shown here that the
three weakest \ion{Th}{2} lines will be undetectable in stars with
much smaller thorium abundances; the spectrum of CS~22892-052
remains the best ``laboratory'' for n-capture studies in metal-poor stars.

Even with \ion{Th}{2} line detection(s), derivation of accurate
thorium abundances requires careful consideration of data and
analysis assumptions.
The transition probabilities of \ion{Th}{2} transitions appear to be 
reasonably well determined.
The standard 4019.12~\AA\ line has received the most attention
in laboratory studies.
Lawler et al. (1990), drawing from the data of Simonsen et al. (1988),
suggest that the log~gf value of this transition has an uncertainty
of only $\pm$0.04~dex.
The more \ion{Th}{2} lines (with accurate gf values) that are 
identified in stellar spectra, the less such transition 
probability uncertainties will matter in the overall error
budget for thorium chronometry; the averaging of abundances from
several lines would certainly help here.

Continuum placement errors and line contamination probably contribute
another $\pm$0.02~dex to the total thorium abundance errors.
Many metal-poor stars now have been observed with very high resolution
and S/N spectra, and the general line weakness of these stars yields
relatively well-defined continua even in the blue-violet spectral
regions where the \ion{Th}{2} lines occur.
But the known blending of the usually-employed 4019~\AA\ line (discussed
above), and the rudimentary current knowledge of blending for the
other thorium lines are the limiting factors here.
Again, the best way to deal with continuum and line contamination
problems is to analyze stars with the largest n-capture overabundances.

Little if any investigation has been made of the effects on thorium
abundances of various stellar atmospheric uncertainties.
The use of different model atmosphere grids (e.g., MARCS, Gustafsson et al.
1975; ATLAS, Kurucz 1992, 1999) certainly alters the derived [Th/H] values,
but does so equally to the abundances of almost all of the usual 
comparison elements (europium, dysprosium, etc.), since transitions
of all of these arise from low excitation levels of their ionized species.
And to our knowledge there are no statistical equilibrium studies
that drop the simplifying assumption of LTE for any of the elements 
considered here. 
Such studies should be attempted, but the atomic data for the elements
considered here simply may be inadequate at present for meaningful insights 
into the magnitude of the abundance changes that could occur.

Finally, even if the n-capture abundances of CS~22892-052 and HD~115444
are in excellent agreement with each other, and the derived [Th/Eu]
abundances also agree well, how do we know that these abundance
ratios represent the typical values for Galactic halo material?
If the [Th/Eu] values have significant star-to-star scatter, then 
they cannot be interpreted in terms of stellar ages (e.g. Goriely
\& Clerbaux 1999).
Resolution of this question is simple in principle, yet it will 
require the derivation
of detailed n-capture abundance patterns in many metal-poor stars.
If the derived [Th/Eu] ratios of a large set of very metal-poor stars
turn out to be all the same within observational errors, then the 
observed thorium abundances undoubtedly are the radioactive
remnants of the thorium that the early Galaxy synthesized --- with
identical but higher [Th/Eu].
On the other hand, if the observed [Th/Eu] ratios end up having significant
scatter among very metal-poor stars, then thorium will lose its
attractiveness for stellar cosmochronometry.
Only future n-capture abundance studies can adequately address this
question.

The situation is less promising for uranium.
The one strong \ion{U}{2} line in a conveniently observable spectral
region lies at 3859.5~\AA, and thus its presence in cool stars is 
normally swamped by CN molecular absorption.
Moreover, its long-lived isotope $^{\rm 238}$U has a half-life of 4.468~Gyrs,
meaning that even if in the early Galaxy uranium was created in reasonable 
amounts compared to thorium (as predicted by {\it r}-process theoretical 
calculations) then its abundance will have been greatly attenuated 
after the 12--15~Gyr history of the Galaxy.
In the most favorable case of CS~22892-052, an upper limit to
the uranium abundance could be derived that was of interest;
do not expect the same meaningful abundance limit in many other stars.

\subsection{Transforming Thorium Abundances into Age Estimates}

From the observed thorium abundances in a metal-poor halo star 
we estimate the ``age'' of its n-capture material
via a simple procedure.
An underlying assumption underlies this calculation: that the 
abundances of the n-capture 
elements observed in the most metal-poor of these stars match the 
solar-system {\it r}-process abundances. 
Observational evidence supporting this assumption has been reviewed in \S2.
The age is derived by comparing the observed stellar abundance ratio 
N$_{\rm Th}$/N$_{\rm Eu}$\footnote{ 
Alternatively, one could consider N$_{\rm Th}$/N$_{\rm Dy}$, or 
N$_{\rm Th}$/N$_{\rm Pt}$, or indeed the abundance ratio of thorium to 
any other predominantly {\it r}-process element (to minimize synthesis 
uncertainties).
However, europium is the most commonly employed stable {\it r}-process
comparison element because its spectral features are easily observed
in most metal-poor stars.}  
with theoretical estimates of the initial value of 
N$_{\rm Th}$/N$_{\rm Eu}$ at the time of formation of these elements 
in an {\it r}-process synthesis event.
Since the radioactive element thorium has a known decay half-life,
the difference in the abundance ratios gives a direct age estimate of 
the stars containing the n-capture material.
For both CS~22892-052 and HD~115444, thorium decay ages of about 14--16~Gyr. 
Repeating the calculation for uranium, the lack of a detectable 
\ion{U}{2} line in CS~22892-052 suggests an age of greater than 11~Gyr.
 
The primary theoretical factor affecting these age estimates involves
uncertain nuclear physics.
Nuclei involved in the {\it r}-process are far from $\beta$-stability 
and therefore the necessary nuclear data for
reliable {\it r}-process predictions are in most cases not
obtainable by experimental determination.
There have, however, been major recent advances in theoretical prescriptions 
for very neutron-rich nuclear data that have been strengthened by recent 
experimental results for very neutron-rich isotopes (see Cowan et al. 1999 
and references therein for a discussion of this.)
One assessment of the reliability of any nuclear mass formulae is
how well the abundance predictions match the observed, stable solar
system {\it r}-process nuclei.
Cowan et al. (1999) found that the best agreement with solar system 
{\it r}-process abundances -- with deviations typically in the 10\%--20\%  
range -- was obtained when they employed the nuclear mass predictions 
from an extended Thomas Fermi model with quenched shell effects 
far from stability (i.e., ``ETFSI-Q'', Pearson, Nayak, \& Goriely 1996).

There is an additional level of uncertainty involved in the zero-decay 
age abundance predictions of the chronometers thorium and uranium, simply 
due to their radioactive nature.
In other words, while the stable heavy solar system elements, such as Pt,
can be compared directly with theoretical predictions, no such
comparison is possible with thorium or uranium.
However, their decay is responsible for the production of the stable lead
and bismuth isotopes, which can be observed and thus compared with
theoretical estimates.
Based upon a number of tests, examining the nuclear physics input which
best reproduced all observables, Cowan et al. (1999) suggested that the 
theoretical errors, when employing a nuclear mass models that are reliable 
far from stability, are of the order of 10-20\%,
resulting in theoretical age uncertainties of $\simeq$ 3-4 Gyr.
Reducing that uncertainty further will require great theoretical and 
experimental nuclear physics efforts to determine
reliable values of the properties of nuclei far from stability.

A final, more encouraging comment is warranted.  
The derivation of ages of metal-rich disk stars from their thorium
abundances is made more complicated by the multiple generations of
n-capture producers that were necessary to build these elements to
their present relatively high abundances.
This means that the radioactive elements like thorium and uranium have multiple
clocks ticking: individual atoms of these elements that exist in 
metal-rich stars may have been synthesized near the start of the
Galaxy, or relatively recently.
There is a confusion of decay times at work.
This source of age uncertainty does not apply to very metal-poor halo
stars.
All of their elements were produced in the first wave of Galactic
nucleosynthesis, and effectively only one clock is at work for the 
whole ensemble of radioactive atoms observed in very metal-poor stars.

\section{Conclusions}

Thorium cosmochronometry is a promising method to derive
the age of our Galaxy, and application of it to the spectra of
very metal-poor halo stars has recently yielded ages that are
in good agreement with determinations from other methods (e.g., Pont et al.
1998; Riess et al. 1998; Perlmutter et al. 1999).
Reliable abundances of both thorium and a number of other n-capture
elements are now available for a handful of very metal-poor stars.
Unfortunately, uncertainties in the age estimates for each of these 
cases remains large, of order $\pm$4~Gyr.
Improvement in this situation depends on a number of factors.
First, it is very important to continue probing the Galactic halo
population for more examples of very metal-poor yet n-capture-rich stars.
Second, for those 30 or so metal-poor stars now known to possess
[n-capture/Fe]~$\gtrsim$~+0.5, more complete n-capture abundance
studies (obviously including very careful attention to thorium)
must be carried out.
Complementary theoretical investigations of {\it r}-process synthesis
must be carried out, for the crucial question in deriving
a thorium-based age of a star is not ``what is the observed [Th/Eu] ratio'',
but rather ``what was the time-zero [Th/Eu]''?

\acknowledgements We are very grateful to our collaborators in
the various n-capture abundance studies that form the basis
of this short review.
Financial support for this research has been provided by NSF
grants AST-9618364 to CS and AST-9986974 to JJC.



\begin{thebibliography}

\bibitem[Burris et al. {}<2000>]{BPACS99}
Burris, D.~L., Pilachowski, C.~A., Armandroff, T., Sneden, C.,
Roe, H., \& Cowan, J.~J. 2000, ApJ, in press

\bibitem[Butcher {}<1987>]{Bu87}
Butcher, H.~R. 1987, Nature, {328}, 127

\bibitem[Cowan et al.{}<1999>]{Cetal99}
Cowan, J.~J., Pfeiffer, B., Kratz, K.-L., Thielemann, F.-K., Sneden, C.,
Burles, S., Tytler, D., \& Beers, T.~C. 1999, ApJ, {521}, 194

\bibitem[Fran\c{c}ois et al. {}<1993>]{FSS93}
Fran\c{c}ois, P., Spite, M., \& Spite, F. 1993, A\&Ap, {274}, 821

\bibitem[Goriely \& Arnould {}<1997>]{GA97}
Goriely, S., \& Clerbaux, B. 1999, A\& A, {346}, 798

\bibitem[Gustafsson et al. {}<1975>]{GBEN75}
Gustafsson B., Bell R.~A., Eriksson, K., Nordlund \AA, 1975, A\&Ap, {42}, 407

\bibitem[Kurucz {}<1992>]{Ku92}
Kurucz, R.~L. 1992, private communication
 
\bibitem[Kurucz {}<1999>]{Ku99}
Kurucz, R.~L. 1999, http://cfaku5.harvard.edu

\bibitem[Lawler et al.{}<1990>]{La90} 
Lawler, J.~E., Whaling, W., \& Grevesse, N. 1990, Nature, {346}, 635

\bibitem[Morell et al. {}<1992>]{MKB92}
Morell, O., K{\"a}llander, D., \& Butcher, H.~R. 1992, A\&Ap, {259}, 543

\bibitem[Norris et al. {}<1997a>]{NRB97a}
Norris, J.~E., Ryan, S.~G., \& Beers, T.~C., 1997a, ApJ, {488}, 350

\bibitem[Norris et al. {}<1997b>]{NRB97b}
Norris, J.~E., Ryan, S.~G., \& Beers, T.~C., 1997b, ApJ, {489}, L169

\bibitem[Pearson et al.{}<1996>]{PNG96}
Pearson, J.~M., Nayak, R.~C., \& Goriely, S. 1996, Phys. Lett. B, {387}, 455

\bibitem[Perlmutter et al. {}<1999>]{Petal99}
Perlmutter, S., Aldering, G., Goldhaber, G., Knop, R. A., Nugent, P.,
Castro, P. G., Deustua, S., Fabbro, S., Goobar, A., Groom, D. E.,
Hook, I. M., Kim, A. G., Kim, M. Y., Lee, J. C., Nunes, N. J., Pain, R.,
Pennypacker, C. R., Quimby, R., Lidman, C., Ellis, R. S., Irwin, M.,
McMahon, R. G., Ruiz-LaPuente, P., Walton, N., Schaefer, B., Boyle, B. J.,
Filippenko, A. V., Matheson, T., Fruchter, A. S., Panagia, N.,
Newberg, H. J. M., \& Couch, W. J. 1999, ApJ, {517}, 565

\bibitem[Pont et al. {}<1998>]{PMTV98}
Pont, F., Mayor, M., Turon, C., VandenBerg, D.~A. 1998, A\&Ap, {329}, 87

\bibitem[Riess et al. {}<1998>]{Retal98}
Riess, A. G., Filippenko, A. V., Challis, P., Clocchiatti, A.,
Diercks, A., Garnavich, P. M., Gilliland, R. L., Hogan, C. J., Jha, S.,
Kirshner, R. P., Leibundgut, B., Phillips, M. M., Reiss, D.,
Schmidt, B. P., Schommer, R. A., Smith, R. C., Spyromilio, J., Stubbs, C.,
Suntzeff, N. B., \& Tonry, J. 1998, AJ, {116}, 1009

\bibitem[Simonsen et al.{}<1988>]{SWSP88}
Simonsen, H., Worm, T., Jessen, P., \& Poulsen, O. 1988, Physica Scripta,
{38}, 370

\bibitem[Sneden et al. {}<2000>]{Setal00}
Sneden, C., Cowan, J.~J., Ivans, I.~I., Fuller, G.~M., Burles, S., 
Beers, T.~C., \& Lawler, J.~E. 2000, ApJ, {467}, 819

\bibitem[Sneden et al. {}<1996>]{SMPCBA96}
Sneden, C.,  McWilliam, A., Preston, G.~W., Cowan, J.~J., Burris, D.~L.,
\& Armosky, B.~J. 1996, ApJ, {467}, 819

\bibitem[Sneden et al. {}<1994>]{SPMS94}
Sneden, C., Preston, G.~W., McWilliam, A. \& Searle, L. 1994, ApJ, {431}, L27

\bibitem[Westin et al. {}<2000>]{WSGC00}
Westin, J., Sneden, C., Gustafsson, B., \& Cowan, J.~J. 2000, ApJ, {530}, 783


\end{thebibliography}
\end{document}